# ASSOCIATIVE MEMORY ON QUTRITS BY MEANS OF QUANTUM ANNEALING

V.E. Zobov[1)], I.S. Pichkovskiy

Kirensky Institute of Physics, Federal Research Center KSC SB RAS, Akademgorodok 50 bld. 38, Krasnoyarsk, 660036 Russia;
Institute of Engineering Physics and Radio Electronics, Siberian Federal University, 28, Kirensky st. Krasnoyarsk, 660074 Russia

We study the functioning of associative memory on three-level quantum elements, qutrites represented by spins with S = 1. The recording of patterns into the superposition of quantum states and their recall are carried out by adiabatic variation of the Hamiltonian with time. To equalize the probabilities of finding the system in different states of superposition, an auxiliary Hamiltonian is proposed, which is turned off at the end of evolution. Simulations were performed on two and three qutrits and an increase in the memory capacity after replacing qubits with qutrits is shown.

At the present stage of development of information technologies, quantum properties of many-body systems open up new possibilities of artificial neural networks [1, 2], including the associative memory [3 - 7]. Quantum superposition of states can be used to store patterns [4 - 8]. As a neuron, a qubit is usually considered - a quantum system with two states, for example, spin with $S = 1/2$. However, in this role a system with three states — qutrit [9–13] can be used. Among the advantages of qutrits over qubits they expect a faster growth of the Hilbert space, and hence the size of the computational basis with the addition of new elements. This circumstance may be important for increasing the memory capacity.

As qurits, it is suggested to use, for example, objects with spin S = 1 in magnetic and crystal fields. These include quadrupole nuclei [9, 12, 13] of deuterium, nitrogen or lithium, as well as NV centers in diamond (paramagnetic color centers formed by an electron on vacancies near the nitrogen atom) [14, 15]. The latter variant is preferable because of the presence of a strong dipole-dipole interaction between NV centers, which is necessary for the implementation of conditional operations in quantum algorithms. Moreover, in this case, there is a large difference in the frequency of transitions between different energy levels, which makes it possible to control the state of the system with the help of transition-selective high-frequency field pulses.

In this paper, we consider the associative memory on the qutrits represented by the spins with $S =1$. The recording of states is accomplished with the help of projection operators. This choice is due to the need to operate with a state with a zero spin projection, the interaction of which with the magnetic field vanishes. The recording and recall of the patterns will be carried out by adiabatic variation of the Hamiltonian over time (by quantum annealing) [4 - 7]. We have proposed a new scheme for alignment of the amplitudes of states in quantum superposition. A numerical simulation of the memory models on two and three qutrits will be performed with a different number of stored patterns.

To implement an associative memory, memorized patterns should be encoded in the states of a quantum system. Take a system of *n* spins with $S =1$, and as a computational basis, we will

---
[1)] e-mail: *rsa@iph.krasn.ru*

use a basis $|m_1, m_2, ..., m_n\rangle$ of the eigenfunctions of the operators $S_i^z$ of the projections of the spins on the Z axis. Each of the projections $m_i$ can take one of three values: 1, 0, -1. Let it is necessary to keep in memory $p$ elements represented by $p$ quantum vectors $|\psi_\mu\rangle = |m_1^\mu, m_2^\mu, ..., m_n^\mu\rangle$, where $\mu = 1, 2, ..., p$. Choose the projection method of memorization [4, 5], in which the recording is performed through the memory Hamiltonian

$$H_{mem} = -\sum_{\mu=1}^{p} |\psi_\mu\rangle\langle\psi_\mu|, \tag{1}$$

where $|\psi_\mu\rangle\langle\psi_\mu|$ is the projector on vector $|\psi_\mu\rangle$. With the help of Hamiltonian (1) we prepare a superposition of memorized states:

$$|\Psi\rangle = \sum_{\mu=1}^{p} a_\mu |\psi_\mu\rangle, \tag{2}$$

where $a_\mu = a(m_1^\mu, m_2^\mu, ..., m_n^\mu)$ are the amplitudes of the states. For this purpose, let us implement the evolution of the system over time $T$ with sufficiently slow (adiabatic) change in time of the Hamiltonian

$$H(t) = (1 - t/T)H_0 + (t/T)H_p, \quad 0 \le t \le T. \tag{3}$$

where $H_p = H_{mem}$, and $H_0$ is the initial Hamiltonian, the ground state of which is easy to prepare. As the last we take an interaction with a magnetic field directed along the $x$ axis:

$$H_0 = -h\sum_{i=1}^{n} S_i^x, \tag{4}$$

the ground state of which is a direct product of the eigenvectors $|1\rangle_x = (|1\rangle + \sqrt{2}|0\rangle + |-1\rangle)/2$ of the spin operators $S_x$ of individual spins with positive eigenvalues 1:

$$|\Psi_0\rangle = |1\rangle_x^{\otimes n}. \tag{5}$$

As a result of the evolution of the system from the initial state (5) under the action of the Hamiltonian (3), we obtain

$$|\Psi(T)\rangle = \hat{Q}\exp\left(-i\int_0^T H(t)dt\right)|\Psi_0\rangle, \tag{6}$$

where $\hat{Q}$ is the ordering operator in time. If the evolution is adiabatic, then the system enters the ground state (2) of the Hamiltonian (1).

In the adiabatic method, the recall of any of the memorized states $|\psi_{prob}\rangle$ is realized by adding a probe Hamiltonian of hint $H_{prob}$ to the target Hamiltonian:

$$H_p = H_{mem} + \Gamma H_{prob}, \tag{7}$$

where $\Gamma$ is a small coefficient that specifies the value of the hint. In [4 - 6], the probe Hamiltonian for spins $S = 1/2$ was set using interaction with a magnetic field. For spins $S = 1$, this option is inconvenient, because it does not allow to take into account the state with zero spin projection. Therefore, we will take

$$H_{prob} = -|\psi_{prob}\rangle\langle\psi_{prob}|. \tag{8}$$

As an example, consider an associative memory on a two-spin system with stored states $|\psi_1\rangle=|0,1\rangle$, $|\psi_2\rangle=|1,0\rangle$ and $|\psi_3\rangle=|-1,-1\rangle$ that corresponds to the Hamiltonian:

$$H_{mem} = -|0,1\rangle\langle 0,1| - |1,0\rangle\langle 1,0| - |-1,-1\rangle\langle -1,-1|. \tag{9}$$

Express the projection operators through the spin operators of an individual spin $i$ ($i=1, 2$):

$$|-1\rangle\langle -1|_i = -S_i^z \frac{1-S_i^z}{2}, \quad |0\rangle\langle 0|_i = 1-(S_i^z)^2, \quad |1\rangle\langle 1|_i = S_i^z \frac{1+S_i^z}{2}, \tag{10}$$

$$|m_1,m_2\rangle\langle m_1,m_2| = |m_1\rangle\langle m_1| \otimes |m_2\rangle\langle m_2| = |m_1\rangle\langle m_1|_1 |m_2\rangle\langle m_2|_2.$$

To calculate the evolution of system (6) with a time-dependent Hamiltonian, we divide the total time $T$ into $N$ small intervals $\Delta t = 0.1$, at each of which we will neglect the change in Hamiltonian (3) [5, 16]. The solution to our problem $|\Psi\rangle$ will be sought as a product:

$$|\psi(t=T)\rangle \cong \prod_{l=0}^{N} \exp\left\{-i\Delta t\left(\frac{l}{N}H_p + \left(1-\frac{l}{N}\right)H_0\right)\right\}|\psi(t=0)\rangle. \tag{11}$$

The calculation results are shown in Fig.1. First, the figure shows that when $\Gamma=0$, we get a superposition (2) with different amplitudes $a_\mu$. This feature of the adiabatic preparation of a superposition state was discussed in theoretical works [7, 8] and was observed experimentally [17]. The reason is that each of the three states approaches the final state at time $T$ along different paths, as can be seen in Fig. 1a for instant energy levels. When $t$ approaches $T$, the energy difference tends to zero and the transition from adiabatic to diabatic evolution occurs [7, 8]. A description of the dynamics of the system at $t \to T$ is given in Supplementary Material [22]. For the probability of finding the system when $t=T$ in a state $|-1,-1\rangle$ using the perturbation theory [4, 7, 8, 18], we obtained the expression

$$|a_3|^2 \approx A(h/T)^{2/3}, \tag{12}$$

which, at $A = 1.22$, is consistent with the results of numerical simulation and qualitatively correctly conveys dependencies on the parameters $h$ from 0.5 to 4 and $T$ from 100 to 400. For equal probabilities of the system being in two other states, $(1-|a_3|^2)/2$ is obtained.

With the inclusion of a probe Hamiltonian in (7), as can be seen in Fig. 1 (b, c), as $\Gamma$ increases, the probability of finding the selected state increases rapidly, while the other two come down. When $\Gamma \ll 1$ in the Supplementary Material [22], the corresponding dependencies are calculated and obtained for them:

at $\Gamma H_{prob} = -\Gamma|0,1\rangle\langle 0,1|$

$$|a_1(\Gamma)|^2 \approx \left(1 + 2\Gamma K_1(T/h)^{2/3}\right)/2, \tag{13}$$

at $\Gamma H_{prob} = -\Gamma|-1,-1\rangle\langle -1,-1|$

$$|a_3(\Gamma)|^2 \approx |a_3(\Gamma=0)|^2\left(1 + 2\Gamma K_3(T/h)^{2/3}\right), \tag{14}$$

where $|a_3(\Gamma=0)|^2$ defined in (12), $K_1=0.203$, $K_3=0.53$. With a further increase in $\Gamma$, the growth of probabilities is accelerated. This is due to the removal of the degeneration between the three states and the transition from the diabatic to the adiabatic evolution as $t$ approaches $T$, accompanied by an increase in the gap $\Gamma$ and a decrease in the probabilities of transitions from the ground state highlighted by the hint. The steepness of dependence on $\Gamma$ increases with increasing $T$ and decreasing $h$.

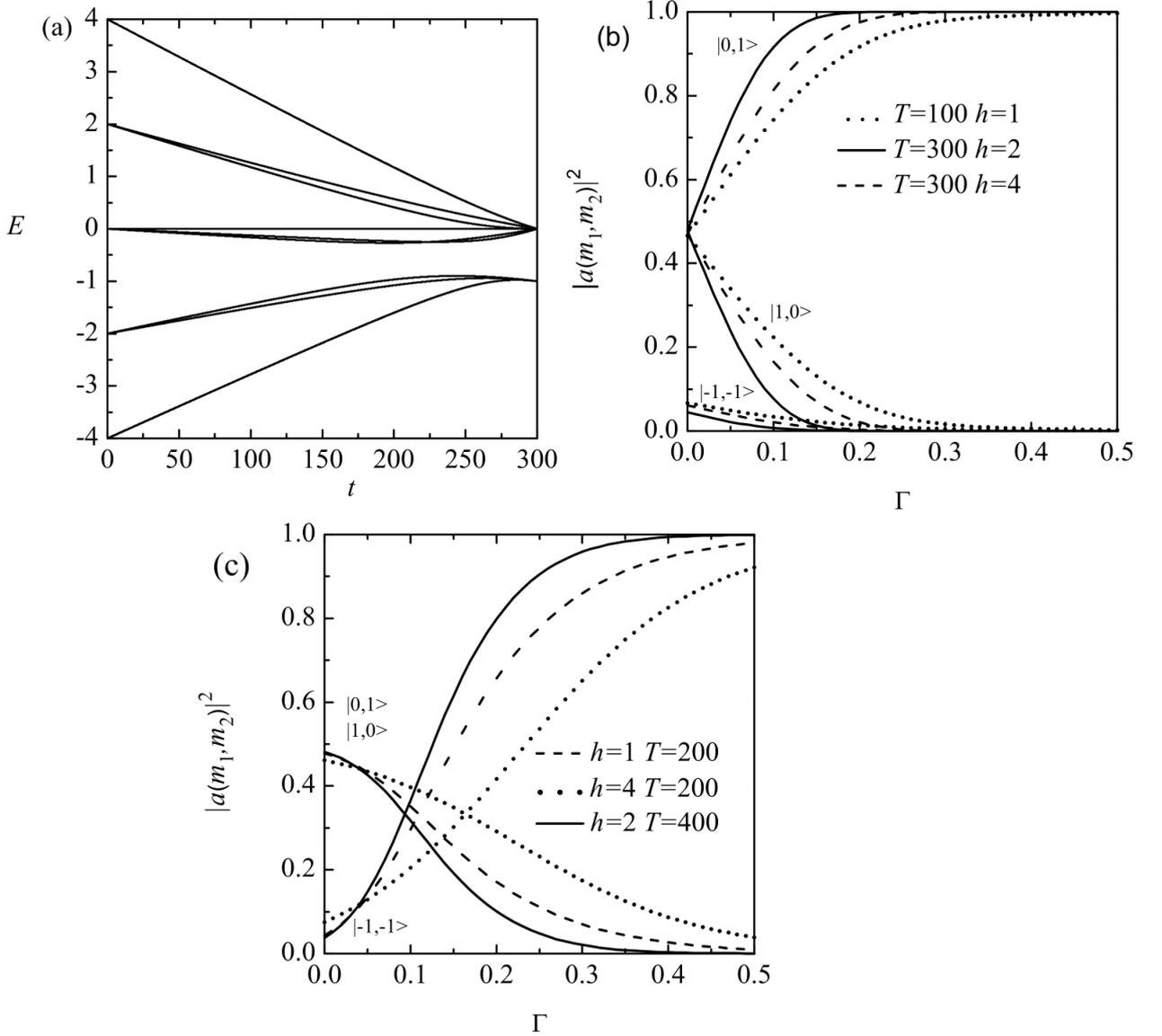

Fig.1. Associative memory on two qutrits with three stored states: $|\psi_1\rangle = |0,1\rangle$, $|\psi_2\rangle = |1,0\rangle$ and $|\psi_3\rangle = |-1,-1\rangle$. (a) instantaneous energy levels of the Hamiltonian $H(t)$ (3) with $H_p = H_{mem}$ (9) versus time as a parameter, with $h = 2$, $T = 300$. (b, c) dependencies on $\Gamma$ of the probabilities of finding the system in the states indicated in the figure with $\Gamma H_{prob} = -\Gamma |0,1\rangle\langle 0,1|$ (b) and with $\Gamma H_{prob} = -\Gamma |-1,-1\rangle\langle -1,-1|$ (c), and for different values of the field $h$ and time $T=\Delta t N$ ($\Delta t=0.1$, $N=T/\Delta t$).

To align the amplitudes in superposition (2) in the case of qubits, it was proposed in [7, 8] to introduce auxiliary spins, with the help of which a multi-spin interaction with the coefficients needed for alignment is created in the effective Hamiltonian.

Another way to obtain an equal superposition state follows from the memory Hamiltonian proposed in [4]

$$H'_{mem} = -|\Psi\rangle\langle\Psi|, \qquad (15)$$

in which $|\Psi\rangle$ is defined in (2) with $a_\mu = 1/\sqrt{p}$. For the above example of two qurits, we have

$$H'_{mem} = -(1/3)\{|0,1\rangle + |1,0\rangle + |-1,-1\rangle\}\{\langle 0,1| + \langle 1,0| + \langle -1,-1|\} = (1/3)H_{mem} + (1/3)H_{help},  \qquad (16)$$

where

$$H_{help} = -\{|0,1\rangle\langle 1,0| + |1,0\rangle\langle 0,1| + |-1,-1\rangle\langle 1,0| + |-1,-1\rangle\langle 0,1| + |0,1\rangle\langle -1,-1| + |1,0\rangle\langle -1,-1|\}, \qquad (17)$$

and $H_{mem}$ is defined in (9). The off-diagonal projectors in (17) will be expressed through the spin projection operators of the individual spin:

$$|0\rangle\langle 1| = S^- S^z, \quad |1\rangle\langle 0| = S^z S^+, \quad |-1\rangle\langle 0| = -S^z S^-, \qquad (18)$$

$$|0\rangle\langle -1| = -S^+ S^z, \quad |-1\rangle\langle 1| = S^- S^-, \quad |1\rangle\langle -1| = S^+ S^+,$$

where we entered the operators:

$$S^\pm = (S^x \pm iS^y)/\sqrt{2}. \qquad (19)$$

The calculations performed with the Hamiltonian (16) showed (see Fig. 2) that at the end of evolution a superposition state is actually realized. However, after adding the probe Hamiltonian of the hint (8), the call to the desired state occurs slowly, i.e. for substantially larger values of $\Gamma$ than in the previous case. This is due to the changed view of the Hamiltonian, the ground state of which we are preparing. In the previous case, there was a choice between three degenerate states, whereas in the latter case, a rearrangement of the only non-degenerate ground state occurs.

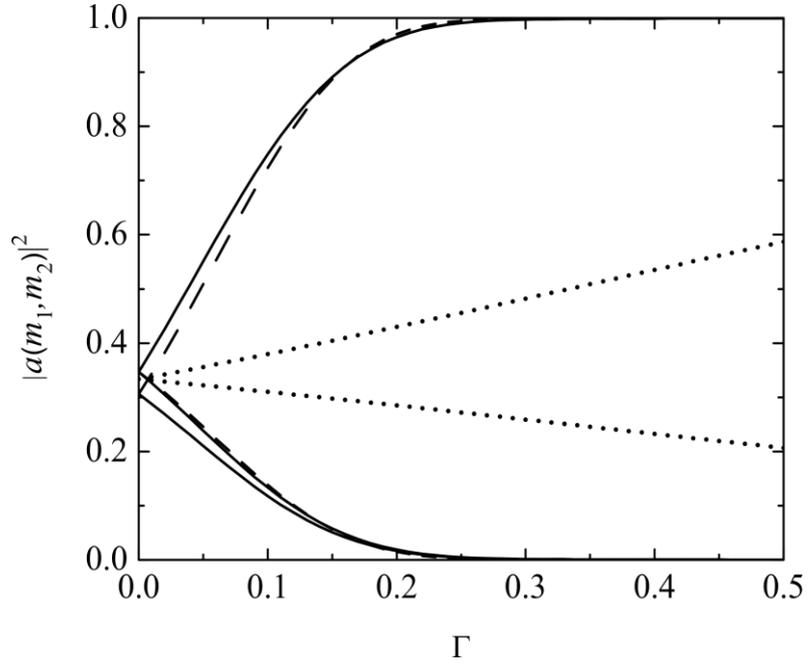

Fig.2. The probabilities of finding the system in three stored states: $|\psi_1\rangle = |0,1\rangle$, $|\psi_2\rangle = |1,0\rangle$ and $|\psi_3\rangle = |-1,-1\rangle$, versus $\Gamma$ with a permanent auxiliary Hamiltonian (16) (dotted lines) or switched off (20) (solid line $\Gamma H_{prob} = -\Gamma|0,1\rangle\langle 0,1|$ and dashed lines $\Gamma H_{prob} = -\Gamma|-1,-1\rangle\langle -1,-1|$). The probability of a state of the same name as a hint deviates upward, and the probabilities of the other two states deviate downward. Values of other parameters are: $h = 2$, $T = 300$.

To combine the merits of the two approaches, we apply the technique proposed in [19]. We will turn off the interaction $H_{help}$ at the end of evolution, i.e. instead of (3), we take the time-dependent Hamiltonian in the form

$$H(t) = (1-t/T)H_0 + (t/T)(1-t/T)H_{help} + (t/T)H_p. \tag{20}$$

The calculation results are presented in Fig.2. From comparison with fig. 1 (b, c) it can be seen that the amplitudes of the three states in the superposition (2) (although not complete) and the type of their dependences on the magnitude of $\Gamma$ really equalized. At the same time, as expected, good sensitivity to the hint remained.

To demonstrate the selectivity of the system's response to the recall of the memorized and unmemorized states, Fig. 3 shows the calculated dependences on $\Gamma$ of the probability of the system being in different states when a probe Hamiltonian is presented on a state that is not in memory: $\Gamma H_{prob} = -\Gamma |1,1\rangle\langle 1,1|$. Now the probability of the recalled state appears at large values of $\Gamma$, which exceed the energy gap between the recorded states and all the others. With such values of $\Gamma$, the called state becomes the ground state [6].

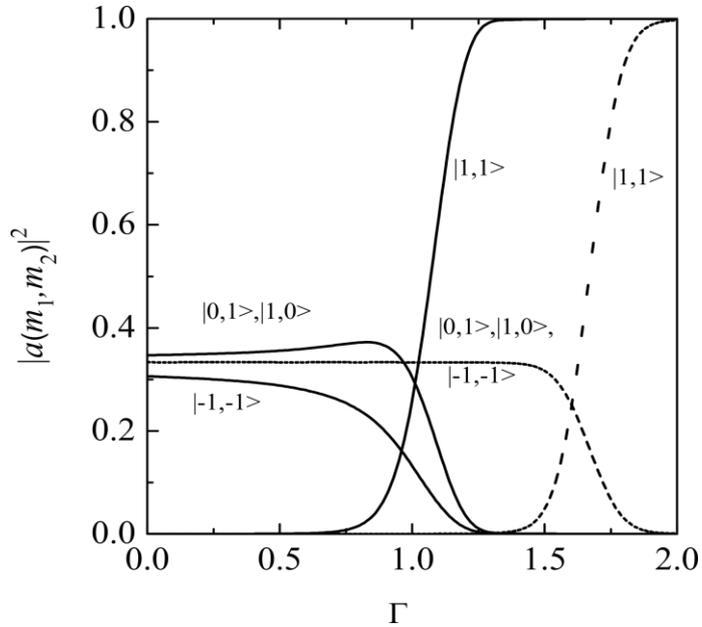

Fig.3. Dependencies of the probabilities of finding the system in the states shown in the figure on the magnitude of the probe Hamiltonian $\Gamma H_{prob} = -\Gamma |1,1\rangle\langle 1,1|$ with the state that is absent in the memory. The dashed lines correspond to the permanently acting auxiliary Hamiltonian (16), the solid lines - to the switched off one (20). Values of other parameters are: $h = 2$, $T = 300$.

The calculations described above were carried out by us for an associative memory with the other three states, as well as with four states. Similar results were obtained. This shows that the memory capacity on two qutrits is higher than on two qubits [4].

Let us turn to a system of three qurits, three spins $S = 1$. The effective Hamiltonians were constructed according to the rules described above for the five memorized states: $|-1,-1,1\rangle$, $|0,0,-1\rangle$, $|1,-1,0\rangle$, $|1,1,1\rangle$ и $|-1,1,-1\rangle$. This time a probe Hamiltonian was chosen:

$$\Gamma H_{prob} = -\Gamma |1,1,1\rangle\langle 1,1,1|. \tag{21}$$

The calculation results are shown in Fig.4. We are convinced of the successful realization of the memorization of the five states on three qutrits. The question of the capacity of associative memory on a larger number of qutrits will be the subject of further research.

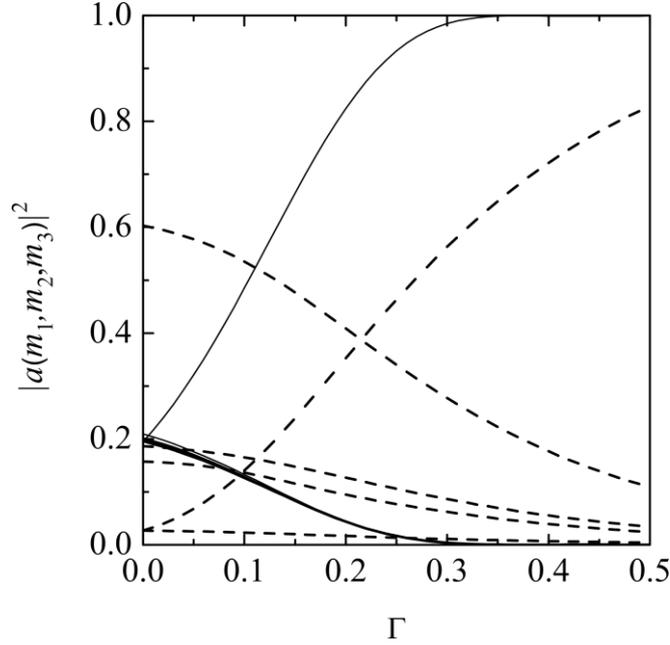

Fig.4. Depending on $\Gamma$, the probabilities of finding the system of three qutrites in five stored states: $|-1,-1,1\rangle$, $|0,0,-1\rangle$, $|1,-1,0\rangle$, $|1,1,1\rangle$ and $|-1,1,-1\rangle$, with the auxiliary switchable (20) Hamiltonian (17) (solid lines) and without one (3) (dashed lines). The probe Hamiltonian of hint is $\Gamma H_{prob} = -\Gamma|1,1,1\rangle\langle1,1,1|$. The probability of a state of the same name as a hint deviates upward, the probabilities of other states deviate downward (dashed lines from top to bottom correspond to states $|0,0,-1\rangle$, $|-1,1,-1\rangle$, $|1,-1,0\rangle$ and $|-1,-1,1\rangle$). Values of other parameters are: $h = 2$, $T = 300$.

Thus, the modeling performed demonstrated the possibility of realizing associative memory on qutrites and pattern recognition by means of quantum annealing. As qurits in the experiment, various three-level quantum systems can be used, for example, objects with spin $S = 1$ in the magnetic and crystal fields listed in the introductory part. The Hamiltonian in the form suitable for experimental implementation is obtained after substitution in $H_{mem}$, $H_{help}$ and $H_{prob}$ the products of three projectors expressed in terms of spin operators (10) and (18). Note that along with the two-spin interaction, which has the form of the Ising interaction or the dipole-dipole interaction, three-spin interactions and interactions containing squares of spin operators are required. Methods for obtaining such interactions with the help of rotation operators selective in transitions between the levels of each of the three spins are described in [14, 15, 20, 21].

1. J. Biamonte, P. Wittek, N. Pancotti, P. Rebentrost, N. Wiebe, and S. Lloyd, Nature **549**, 195 (2017).
2. V. Dunjko and H. J. Briegel, Reports on Progress in Physics **81**, 074001 (2018).
3. Robert Callan, The essence of neural networks. (Prentice Hall Europe, Copyright, 1999).
4. R. Neigovzen, J.L. Neves, R. Sollacher, and S.J. Glaser, Phys. Rev. A **79,** 042321 (2009).
5. H. Seddiqi and T.S. Humble, Front. Phys. **2**, 79 (2014).
6. S. Santra, O. Shehab, and R. Balu, Phys. Rev. A **96**, 062330 (2017).
7. C. Dlaska, L.M. Sieberer, and W. Lechner, Phys. Rev. A **99**, 032342 (2019).
8. L. M. Sieberer and W. Lechner, Phys. Rev. A **97**, 052329 (2018).
9. R. Das, A. Mitra, V. Kumar, and A. Kumar, Int. J. Quantum Inf. **1**, 387 (2003).
10. A.B. Klimov, R. Guzmán, J.C. Retamal, and C. Saavedra, Phys. Rev. A **67,** 062313 (2003).
11. B. Tamir, Phys. Rev. A **77**, 022326 (2007).
12. V.E. Zobov and D.I. Pekhterev, JETP Letters **89**, 260 (2009).
13. V.E. Zobov and V.P. Shauro, JETP **113,** 181 (2011).
14. S. Choi, N.Y. Yao, M.D. Lukin, Phys. Rev. Lett. **119,** 183603 (2017).
15. M.F. O'Keeffe, L. Horesh, D.A. Braje, and I.L. Chuang, New J. Phys. **21,** 023015 (2019).
16. M. Steffen, W. van Dam, T. Hogg, G. Breyta, and I. Chuang, Phys. Rev. Lett. **90**, 067903 (2003).
17. S. Mandra, Z. Zhu, and H.G. Katzgraber, Phys. Rev. Lett. **118**, 070502 (2017).
18. L.D Landau and E.M. Lifshitz, Quantum Mechanics: Non-relativistic Theory, 3rd ed. (Pergamon Press, Oxford, 1977).
19. E. Crosson, E. Farhi, C. Yen-Yu Lin, Han-Hsuan Lin, and P. Shor, arXiv:1401.7320.
20. V.E. Zobov and A.S. Ermilov, JETP **114**, 923 (2012).
21. V.E. Zobov and I.S. Pichkovskiy, Proc. SPIE 11022, International Conference on Micro- and Nano-Electronics 2018, 110222V (15 March 2019); doi: 10.1117/12.2521253.
22. See Supplemental Material.

Supplemental material to article "ASSOCIATIVE MEMORY ON QUTRITS BY MEANS OF QUANTUM ANNEALING"

V.E. Zobov, I.S. Pichkovskiy

Let us consider the dynamics of the system of two spins $S=1$ with a Hamiltonian (3) and (9) in the process of transition at $t \to T$ to a threefold degenerate ground state (2). At each time moment $t$ for Hamiltonian (3) we can find eigenstates $|\varphi_k(t)\rangle$ and eigenvalues of energy $E_k(t)$ (see Fig. 1a)

$$H(t)|\varphi_k(t)\rangle = E_k(t)|\varphi_k(t)\rangle \qquad (S.1)$$

$$|\varphi_k\rangle = \sum_{m_1,m_2} a_k(m_1,m_2)|m_1,m_2\rangle \qquad (S.2)$$

At $t = 0$, the system is in the ground state (5) of the Hamiltonian $H_0$ (4). If the sweep of the Hamiltonian is adiabatic ($T \gg 1$), then for most of the time of evolution the system remains in the instantaneous ground state, the form (S.2) of which, naturally, will change. As $t$ approaches $T$, the difference in energy between the three lower energy levels tends to zero. The adiabaticity is violated and the sweep will induce diabatic transitions between these levels under the influence of the perturbation

$$V = -h(1 - t/T)(S_1^x + S_2^x). \qquad (S.3)$$

The six upper levels are still separated by a large energy gap and remain unpopulated.

The perturbation (S.3) has zero matrix elements between the selected stored states $|0,1\rangle$, $|1,0\rangle$ and $|-1,-1\rangle$, therefore the calculation should be carried out in two stages [18]. First, find the corrections to these eigenstates from the transitions connecting them with the upper levels separated by gap 1, and then the transitions between them. At the first stage we find

$$\begin{aligned}|\varphi_1(t)\rangle &= c_1|0,1\rangle + c_2(|1,1\rangle + |0,0\rangle + |-1,1\rangle) \\ |\varphi_2(t)\rangle &= c_1|1,0\rangle + c_2(|1,1\rangle + |1,-1\rangle + |0,0\rangle), \\ |\varphi_3(t)\rangle &= c_3|-1,-1\rangle + c_2(|0,-1\rangle + |-1,0\rangle)\end{aligned} \qquad (S.4)$$

where the coefficients $c_1$, $c_2$ and $c_3$ have the form:

$$c_1 = \sqrt{1 - 3c_2^2}, \quad c_2 = \frac{h}{\sqrt{2}}\left(1 - \frac{t}{T}\right), \quad c_3 = \sqrt{1 - 2c_2^2}. \qquad (S.5)$$

We now calculate the matrix of Hamiltonian (3) in the basis (S.4)

$$H_{eff} = \begin{bmatrix} -c_1^2 t/T - 6c_1 c_2^2 & -4c_1 c_2^2 & -3c_2^3 \\ -4c_1 c_2^2 & -c_1^2 t/T - 6c_1 c_2^2 & -3c_2^3 \\ -3c_2^3 & -3c_2^3 & -c_3^2 t/T - 4c_2^2 c_3 \end{bmatrix}. \qquad (S.6)$$

Thus, we obtained the effective Hamiltonian [7, 8], which drives the dynamics of the three lower states.

It is useful to go to the basis of eigenstates:

$$|\varphi_\pm(t)\rangle = \frac{1}{\sqrt{2}}|\varphi_1(t)\rangle \pm \frac{1}{\sqrt{2}}|\varphi_2(t)\rangle, \quad |\varphi_3(t)\rangle. \qquad (S.7)$$

In the new basis we obtain, retaining in the matrix elements the leading contributions on a small value $c_2$,

$$H_{eff} = \begin{bmatrix} -t/T - 7c_2^2 & 0 & -3\sqrt{2}c_2^3 \\ 0 & -t/T + c_2^2 & 0 \\ -3\sqrt{2}c_2^3 & 0 & -t/T - 2c_2^2 \end{bmatrix}. \tag{S.8}$$

In the approximation considered, the state $|\varphi_-(t)\rangle$ is independent of the other two states. Therefore, we have

$$|\psi(t)\rangle = a_+ |\varphi_+(t)\rangle + a_3 |\varphi_3(t)\rangle, \tag{S.9}$$

and for the amplitudes before eigenstates from the Schrödinger equation we get a system of two equations:

$$\begin{aligned} \frac{da_+}{dt} &= i\left(\frac{t}{T} + 7c_2^2\right)a_+ + i3\sqrt{2}c_2^3 a_3 \\ \frac{da_3}{dt} &= i3\sqrt{2}c_2^3 a_+ + i\left(\frac{t}{T} + 2c_2^2\right)a_3 \end{aligned}. \tag{S.10}$$

Since $c_2 \ll 1$, the off-diagonal terms in (S.10) can be regarded as a small perturbation [18]. In the first order of perturbation we find at $t = T$

$$a_3 = i3\sqrt{2}\int_{t_d}^{T} \exp\left\{i\int_{t_d}^{t} 5c_2^2(t_1)dt_1\right\} c_2^3(t)dt. \tag{S.11}$$

When receiving (S.11) from (S.10), we assumed that at $t = t_d$ $a_3 = 0$, but $a_+ = 1$, i.e. $t_d$ is the time at which the adiabatic dynamics becomes the diabatic one [8]. In the formula (S.11) and in all subsequent ones, we do not write the phase factor due to the average energy of the two states, since it does not affect the values of the state probabilities. Turning to the variable $x=(1-t/T)$ and transforming the integral, we obtain

$$a_3 = i\frac{Th^3}{2\omega^4} \int_0^{(x_d\omega)^3} e^{iz} z^{1/3} dz, \tag{S.12}$$

where $\omega^3 = \frac{5}{6}Th^2$. As a consequence, for the probability of finding the system in the state $|-1,-1\rangle$, we obtain the estimate

$$|a_3|^2 \approx A(h/T)^{2/3}, \tag{S.13}$$

where $A = (1.2)^{8/3}(R_3^2 + I_3^2)/4$, and $R_3, I_3$ are real and imaginary parts of the integral standing in (S.12). The modulus of this integral grows with increasing of $(x_d\omega)^3$, reaches a value of 6 at $(x_d\omega)^3 \cong 4$, and then oscillates. We took the value 3 for it. The corresponding dependences (S.13) (or (12)) on the parameters are shown in Fig. S1 in comparison with the results obtained by numerical calculation of evolution.

Add to Hamiltonian (3) a probe term

$$\Gamma H_{prob} = -\Gamma |-1,-1\rangle\langle -1,-1|. \tag{S.14}$$

The corresponding term $\Gamma$ is added to (S.8) and (S.10), and therefore (S.11) takes the form

$$a_3(\Gamma) = i3\sqrt{2}\int_{t_d}^{T} \exp\left\{i\int_{t_d}^{t}[5c_2^2(t_1) - \Gamma]dt_1\right\} c_2^3(t)dt. \tag{S.15}$$

Assuming the smallness of $\Gamma \ll 1$, we expand the exponential function in (S.15) along $\Gamma$ for the first order, then in the expression (S.12) for $a_3$ to be added the second term

$$\delta a_3(\Gamma) = \frac{Th^3}{2\omega^4} \int_0^{(x_d\omega)^3} e^{iz} z^{2/3} dz \frac{T\Gamma}{\omega},  \tag{S.16}$$

taking which into account we will get instead of (S.13)

$$|a_3(\Gamma)|^2 \approx |a_3(\Gamma=0)|^2 \left(1 + 2K_{4/3} \frac{T\Gamma}{\omega}\right),  \tag{S.17}$$

where $K_{4/3} = \frac{R_3 I_4 - R_4 I_3}{R_3^2 + I_3^2}$, and $R_4, I_4$ are real and imaginary parts of the integral in (S.16). As the value $(x_d\omega)^3$ changes from 0 to 6, the $K_{4/3}$ coefficient increases from 0 to ~0.6, and then begins to oscillate. To clarify its value, we compared it with the results following from the numerical calculations of the evolution. In Fig. S1 we gave the calculated values of the derivative $\frac{d}{d\Gamma}|a_3(\Gamma)|^2$ at $\Gamma \to 0$ and the values following from (S.17) (or (14)) with $K_{4/3}$=0.5.

Now add in Hamiltonian (3) the another probe term

$$\Gamma H_{prob} = -\Gamma|0,1\rangle\langle 0,1|.  \tag{S.18}$$

The effective Hamiltonian (S.8) takes the form

$$H_{eff} = \begin{bmatrix} -t/T - 7c_2^2 - \Gamma/2 & -\Gamma/2 & -3\sqrt{2}c_2^3 \\ -\Gamma/2 & -t/T + c_2^2 - \Gamma/2 & 0 \\ -3\sqrt{2}c_2^3 & 0 & -t/T - 2c_2^2 \end{bmatrix}  \tag{S.19}$$

As in the previous cases, we assume that at $t = t_d$ $a_3 = a_- = 0$, and $a_+(t=t_d)=1$. In the first order of the perturbation theory, we obtain at time $T$

$$\delta a_-(\Gamma) = i\frac{T\Gamma}{6\Omega} \int_0^{(x_d\Omega)^3} e^{iz} \frac{dz}{z^{2/3}},  \tag{S.20}$$

where $\Omega^3 = \frac{4}{3}Th^2$. Under the adiabatic conditions $T \gg 1$ we take the limit value of the integral with $(x_d\Omega)^3 = \infty$: $\Gamma(1/3)[\sqrt{3}/2 - i/2]$, where $\Gamma(1/3)$ - gamma function. Returning to the computational basis, we obtain the result (13), given in the main text. In fig. S1, we compared the coefficient before $\Gamma$ from this dependence with the result for the derivative $\frac{d}{d\Gamma}|a_1(\Gamma)|^2$ at $\Gamma \to 0$, obtained from numerical calculations of evolution. Good agreement is observed.

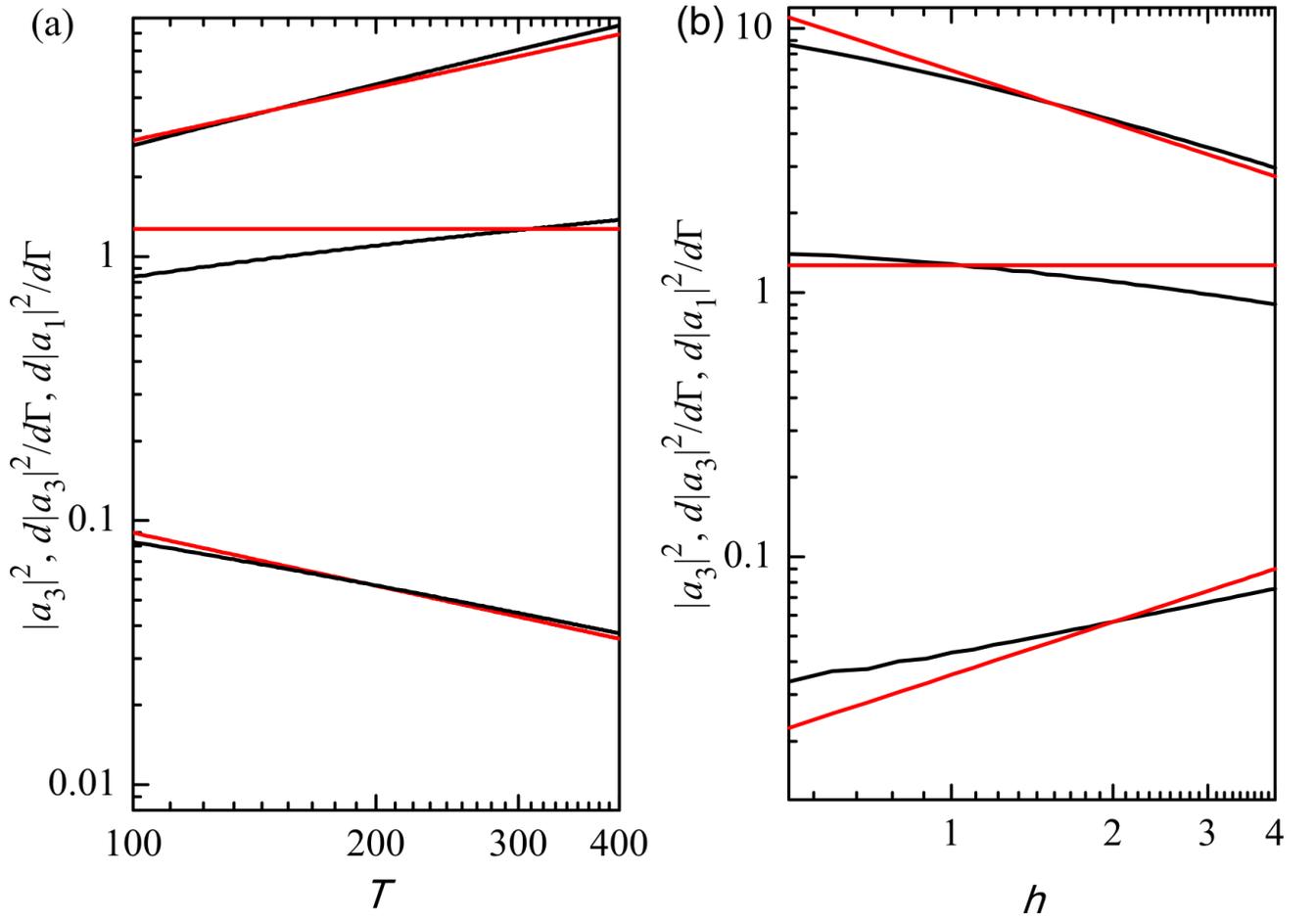

Fig. S1. Comparison between the results of numerical simulation (black lines) with the results of the perturbation theory (red lines) (in logarithmic coordinates). Three pairs of lines correspond, from the bottom to the top: the probability of state $|a_3|^2$, the derivatives with respect to the parameter $\Gamma$ (at $\Gamma \to 0$) $\frac{d}{d\Gamma}|a_3(\Gamma)|^2$, and $\frac{d}{d\Gamma}|a_1(\Gamma)|^2$; (a) as function of the duration of evolution $T$ with $h = 2$, (b) as function of the field strength with $T = 200$.